\documentclass[prl,aps,twocolumn]{revtex4}
\usepackage{graphicx} 
\usepackage[usenames]{color}
\usepackage{amsmath,amssymb}
\usepackage{gensymb}
\usepackage{natbib}
\usepackage{xcolor}
\def\strutdepth{\dp\strutbox}
\def\nw#1{\strut\vadjust{\kern-\strutdepth\vtop to0pt{\vss\hbox to\hsize
{\hskip\hsize\hskip5pt$\leftarrow$\hss\strut}}}{\em #1}}
\usepackage{rotating}

\usepackage{color}
\definecolor{red}{rgb}{1,0,0}

\usepackage{blindtext}

\begin{document}

\title{Ion and site correlations of charge regulating surfaces: a simple and accurate theory}

\author{Martin Trulsson}
\affiliation{Theoretical Chemistry, Lund University, Sweden}

\begin{abstract}
Charge regulation is a fundamental mechanism in most chemical, geochemical, and biochemical systems.
Various minerals surfaces and proteins are well-known to change their charge state as a function of the activity of the hydronium ions, \emph{i.e.,}~the $p$H. Besides being modulated by the p$H$, the charge state is sensitive to salt concentration and composition due to screening and ion correlations. Given the importance of electrostatic interactions, a reliable and straightforward theory of charge regulation would be of utmost importance. This paper presents such a theory that accounts for salt screening, site and ion correlations. Our approach shows an impeccable agreement compared to Monte Carlo simulations and experiments of 1:1 and 2:1 salts. We furthermore disentangle the relative importance of site-site, ion-ion and ion-site correlations. Contrary to previous claims, we find that ion-site correlations are subdominant to the two other correlation terms.
 \end{abstract}

\date{\today}
\maketitle
  
\begin{figure*}[!htbp]
\centering
\includegraphics[width=0.23\textwidth]{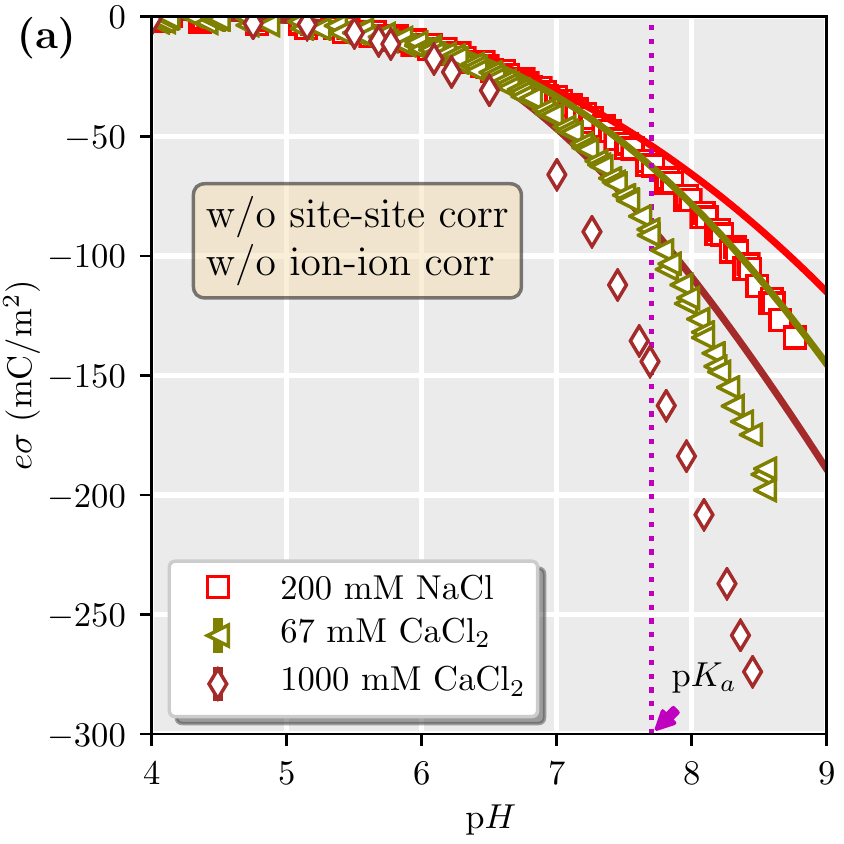}
\includegraphics[width=0.23\textwidth]{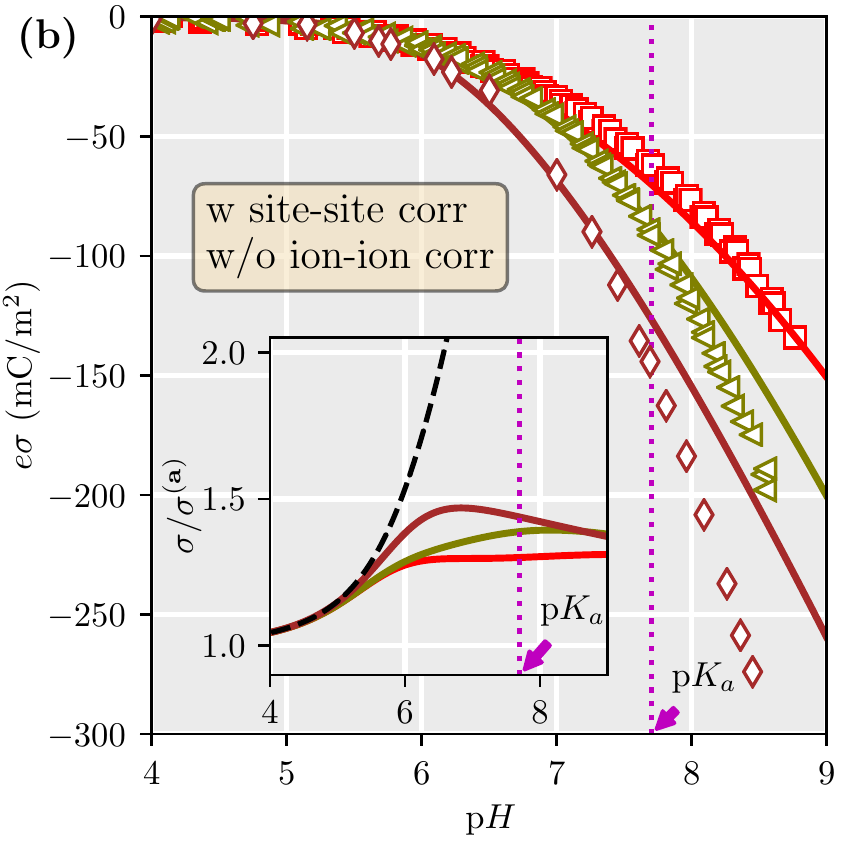}
\includegraphics[width=0.23\textwidth]{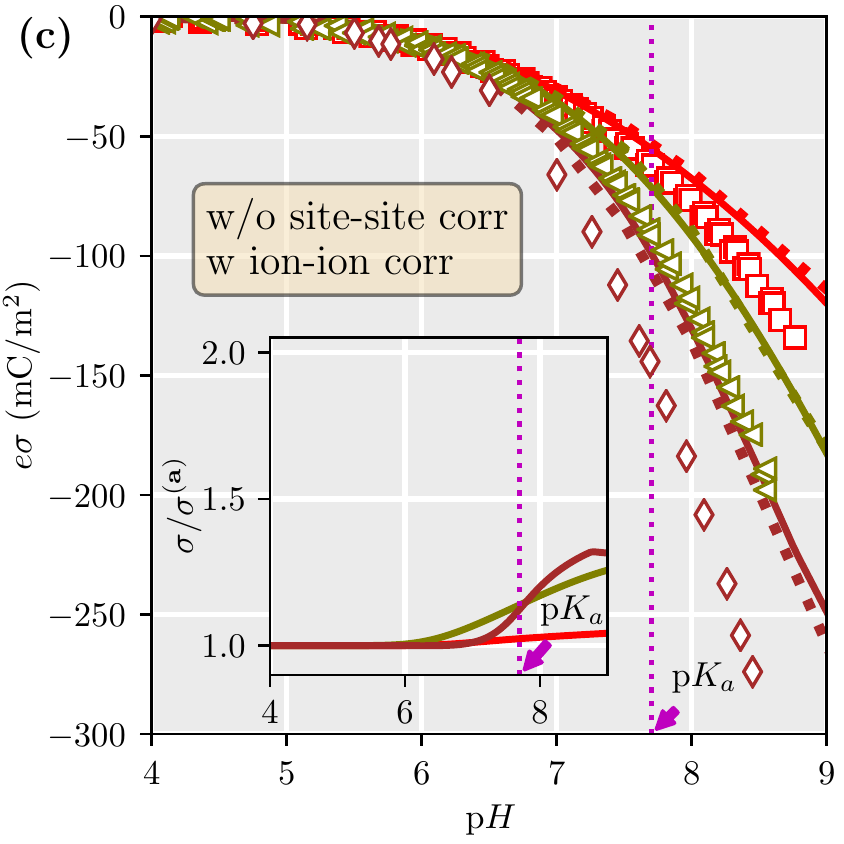}
\includegraphics[width=0.23\textwidth]{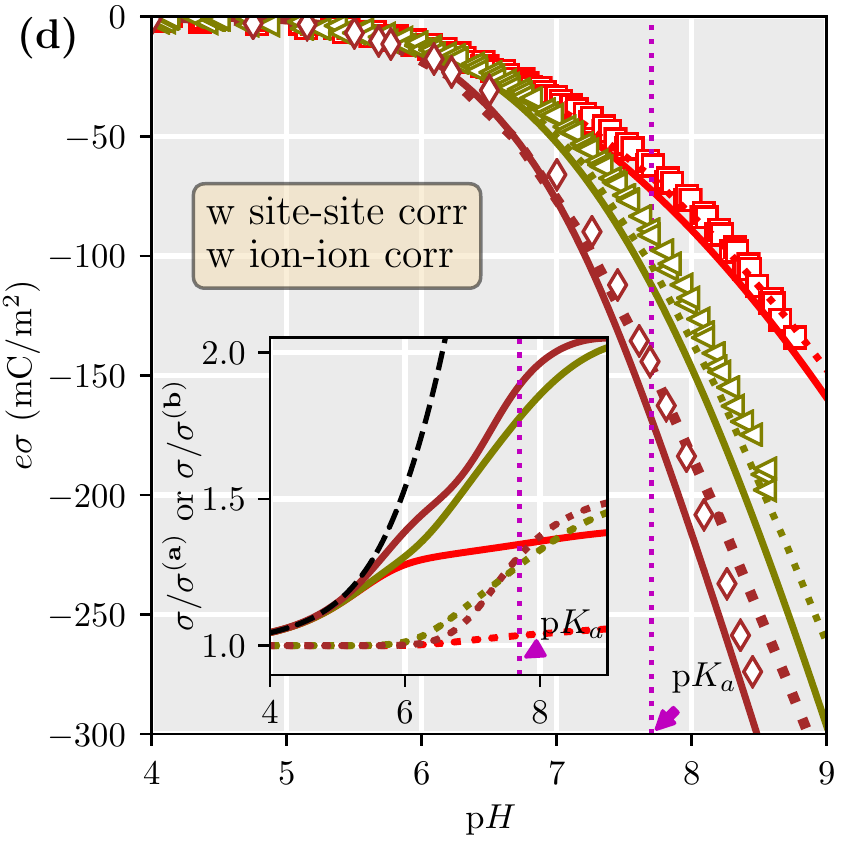}
  \caption{Comparison between the different levels of theory {\bf (a)} simple capacitor {\bf (b)} with site-site correlations, {\bf (c)} with ion-ion correlations and {\bf (d)} with both ion-ion and site-site correlations. Markers are experimental data, solid lines theory and dashed lines MC results. Insets show (solid lines) the relative increase compared to {\bf (a)} by including various terms. The inset of {\bf (d)} also shows (dashed lines) the relative increase compared to having only site-site correlations included (\emph{i.e.,} compared to {\bf (b)}).}
  \label{fgr:Titr21}
\end{figure*}

\begin{figure}[!htbp]
\centering
\includegraphics[width=0.35\textwidth]{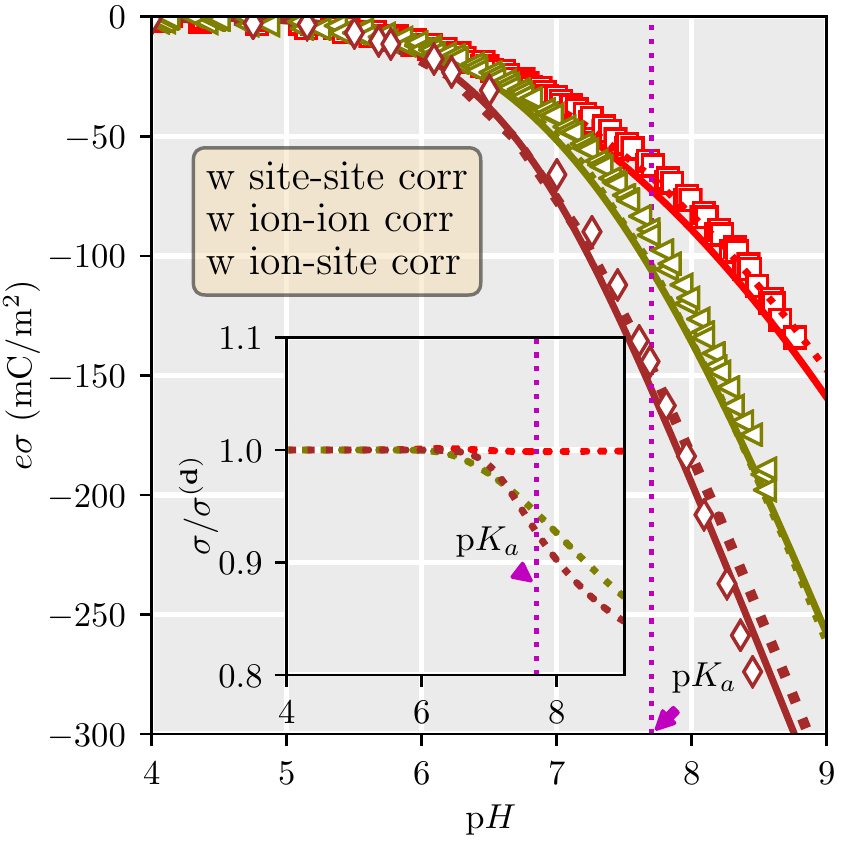}
    \caption{Comparison between the experimental, Monte Carlo, and theory titration curves. The latter includes all (ion-ion, ion-site, and site-site) correlations. Inset shows the relative difference between including ion-site correlations and not (denoted by $\sigma^{\bf (d)}$). }
      \label{fgr:Titr21F}
\end{figure}

Surface dissociation or association of hydronium ions in aqueous solutions is of great importance for many chemical and biological systems. Immersed nanoparticles (\emph{e.g.,}~proteins \cite{Lund05}) or surfaces \cite{Dove05} acquire electrostatic charges regulated by hydronium ion concentration, salt concentration, and many other factors. The electrostatic charge is one key factor for stabilising colloidal suspensions, as they would otherwise aggregate due to attractive van der Waals forces unless other protective interactions are at play (\emph{e.g.,}~sterically repelling polymer brushes). The praised DLVO-theory\cite{Derjaguin41,Verwey48}, including attractive van der Waals forces and repulsive electrostatic forces, helps us  
predict colloidal stability between equal and simple colloidal particles, It successfully predicts a lowered stability at higher salt concentrations as stabilising electrostatic forces are screened out. \\ 
The most straightforward approach to predict a surface's charge, as needed for the DLVO-theory, uses the Henderson-Hasselbalch formula \cite{Po01}, which relates the surface ionisation to the hydronium ion concentration. However, this formula does not account for the electrostatic potential build-up at the surface due to the mutual electrostatic energy between the charged surface and its diffuse counterion layer. Hence, it typically overestimates the surface charge density. To account for this effect, one typically rely on Poisson-Boltzmann calculations \cite{Nimham71,Behrens01,Trefalt16} or Monte Carlo (MC) simulations \cite{Labbez09,Labbez06}, both of which account for both p$H$ and salt concentration dependence \cite{Labbez09}. Unlike MC simulations, Poisson-Boltzmann theory is, however, well-known to be insufficient as soon as the counterions of the charged surface are multivalent (divalent and above) due to lack of ion-ion correlations in the mean-field theory. Moreover, Poisson-Boltzmann calculations generally do not account for the discreteness of the surface sites, even though Bakhshandeh~\emph{et al.}~\cite{Bakhshandeh19} have recently accounted for the latter showing an excellent agreement between theory and MC simulations of charge regulating nanoparticles in the presence of 1:1 salts.\\
Here, we extend this work, accounting for ion correlations, with a straightforward yet highly accurate theory for planar surfaces. 
We compare our theoretical predictions to available experimental data \cite{Dove05} and previous MC simulations \cite{Labbez09}, within the primitive model, both in 1:1 and 2:1 salt solutions and with excellent agreement. Moreover, we disentangle the various contributions to the overall titration behaviour and find that site-site correlations are important for all the studied cases. Ion-ion correlations are, as expected, only relevant at high electrostatic coupling parameters, \emph{i.e.}~when the excess counterion concentration is high close to the surface due to a high surface charge density, and these counterions are multivalent.
In contrast to previous claims \cite{Labbez09}, we find that the ion-site correlations are generally subdominant. \\
\newline
{\it Theory -} Instead of solving the full non-linear Poisson-Boltzmann equation, we rely on just solving the boundary condition, assuming a flat impenetrable titratable surface.
The contact theorem specifies that 
\begin{equation}
2 \pi l_B \sigma^2 = \sum_i \Big(\rho_i^{\rm s} -\rho_{i}^{\rm bulk}\Big)=\sum_i (e^{-\beta z_i e\phi^{\rm ion}}-1)\rho_{i}^{\rm bulk},
\label{contact}
\end{equation}
where $l_B$ is the Bjerrum length, $\sigma$ the surface charge density (in inverse area units), $e$ the elementary charge, $\rho_{i}^{\rm bulk}$ and $\rho_{i}^{\rm s}$ the bulk reservoir and surface densities of species $i$, $z_i$ the valency of species $i$, $\beta\equiv1/k_B T$ the inverse temperature,  and $\phi^{\rm ion}$ the electrostatic potential on an ion at the surface. 
The hydronium activity $a_{\rm H_3O^+}$ is traditionally given in logarithmic units ${\rm p}H=-\log_{10} a_{\rm H_3O^+}$ and the acid equilibrium constant ${\rm p}K_a=-\log_{10} K_a$ of the reaction $-{\rm OH}+{\rm H_2O}(l) \rightleftharpoons -{\rm O^-}+{\rm H_3O^+}(aq.)$. In most cases, one assumes that the activity of the hydronium ion is close to its ideal part, \emph{i.e.,}~its concentration, by $a_{\rm H_3O^+}\approx [{\rm H_3O^+}]$ where $[{\rm H_3O^+}]$ (usually abbreviated by $[{\rm H^+}]$) is the concentration in units of mol/L. $K_a$ is then given in the same units. 
The ionisation degree of the surface charge density $\alpha$, is given by the simple equilibrium relationship
 \begin{figure}[!htbp]
\centering
\includegraphics[width=0.23\textwidth]{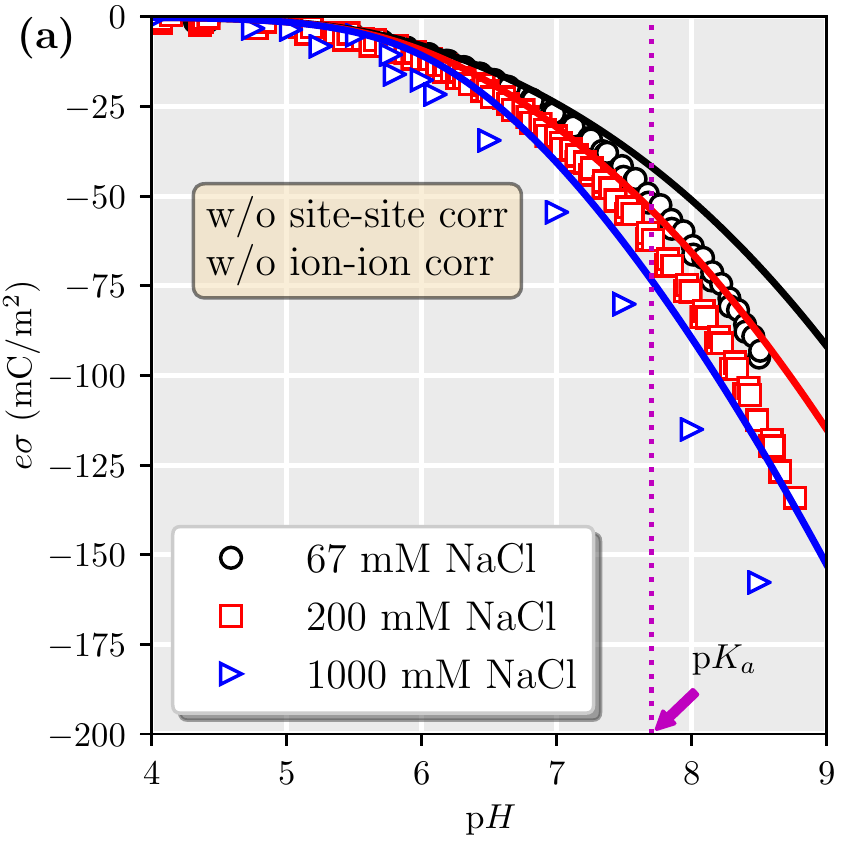}
\includegraphics[width=0.23\textwidth]{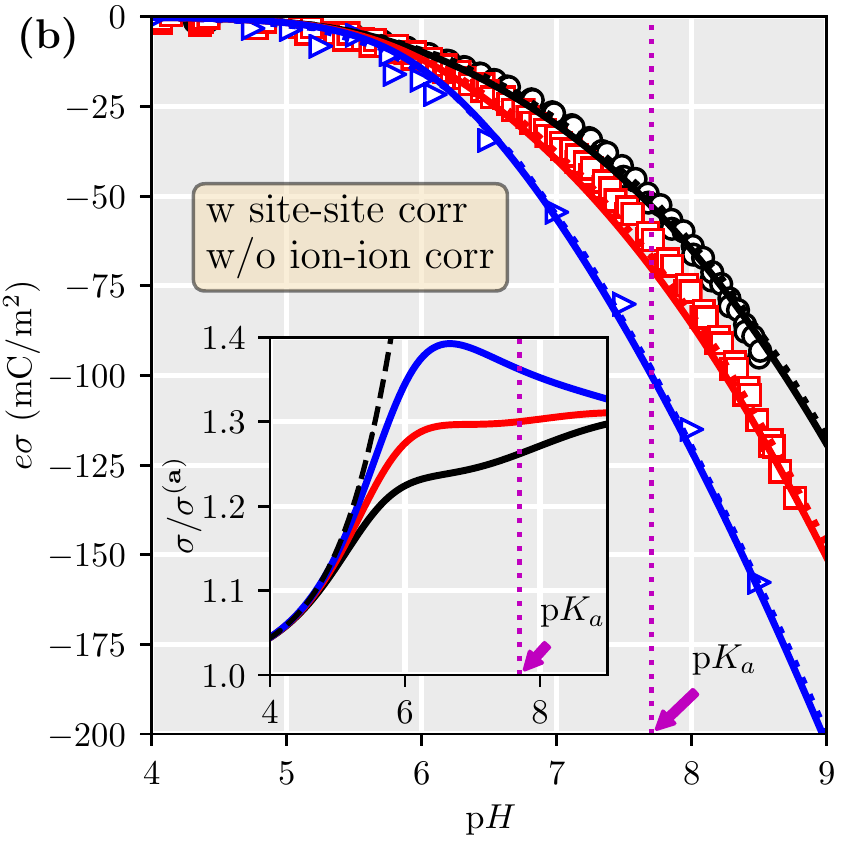}
  \caption{Comparison between the different levels of theory {\bf (a)} simple capacitor {\bf (b)} with site-site correlations. Markers are experimental data \cite{Dove05}, solid lines the presented theory and dashed lines MC results \cite{Labbez09}. Inset shows the relative increase in surface charge density when including site-site correlations compared to w/o (denoted by $\sigma^{\bf (a)})$. The dashed line in the inset shows the prediction according to Eq.~\ref{eq:lowI}.}
  \label{fgr:Titr11}
\end{figure}
\begin{equation}
\beta e \phi^{\rm site}=\ln \Big( \frac{\alpha}{1-\alpha} \Big)+\ln(10) ({\rm p}K_{\rm a}-{\rm p}H),
\label{titr}
\end{equation}
where we assumed negatively charged (acid) surface groups (on the type $\rm -O^-$) and $\phi^{\rm site}$ is the electrostatic potential on a surface group. The surface charge density is equal to  $\sigma = -\alpha \sigma^{\rm site}$, where  $\sigma^{\rm site}$ is the site group density, including both neutral and negative sites.
We decompose the electrostatic potential on salt ion at the surface as $\phi^{\rm ion}=\phi^{\rm mf}+\phi^{\rm ii}+\phi^{\rm ex}$, where $\phi^{\rm mf}$ is the mean-field, \emph{e.g.,}~Poisson-Boltzmann, electrostatic potential at the surface, $\phi^{\rm ii}$ an electrostatic potential correction on the ion due to the ion-ion correlations, and $\phi^{\rm ex}$ an excluded volume correction to the electrostatic potential correction due to the finite size of an ion.
Similarly, one can decompose the site potential as $\phi^{\rm site}=\phi^{\rm mf}+\phi^{\rm cap}+\phi^{\rm ss}$, where $\phi^{\rm cap}$ is the capacitance contribution and $\phi^{\rm ss}$ a correction due to site-site correlations. The capacitance $C_{\rm S}$ is related to the finite closest approach (also denote the Stern layer) between the charges of the ions and that of the surface sites and gives rise to a potential contribution as $\phi^{\rm cap} = e\sigma/C_{\rm S}$, where $C_{\rm S}=\varepsilon_0\varepsilon_r/d_{is}$, $d_{\rm is}$ the closest separation, perpendicular to the surface, between the charges of an ion and a site, and $\varepsilon_0\varepsilon_r$ is the absolute dielectric permittivity.   
We approximate the site-site correlations as 
\begin{equation}
\beta e\phi^{\rm ss}= -\mathrm{sgn}(\sigma) M_{\rm sq} l_B \sqrt{|\sigma|},
\label{eq:ss}
\end{equation}
where $M_{\rm sq}\approx 1.949$ is the Madelung constant for a square lattice \cite{Bonsall77} \footnote{Assuming a hexagonal structure instead renders a similar value, $M_{\rm hex}\approx 1.961$. The present choice is motivated by the 
site structure used in the MC simulations.}, which corresponds to having the charges of the sites at a maximum distance away from each other on a square lattice, 
and  the ion-ion correlations via a first loop correction \cite{Netz00} as
\begin{equation}
\beta e\phi^{\rm ii}= c_0 \Xi/z_{\rm s},
\end{equation}
where $c_0=(\pi/8-0.3104)\approx 0.08230$, $\Xi$ is an average coupling parameter, and $z_{\rm s}$ an average charge of the ions acting as co- and counterions to the charged surface.
We approximate the latter quantity with the average valency of the ions at surface contact  
\begin{equation}
z_{\rm s} = \frac{\sum_i z_i \rho_i^{\rm bulk} e^{-\beta z_i e\phi^{\rm ion}}}{\sum_i  \rho_i^{\rm bulk} e^{-\beta z_i e\phi^{\rm ion} }}.
\label{avgZ}
\end{equation}
Following Refs.~\cite{Trulsson17a, Trulsson17b} we define the average coupling parameter as $\Xi = 2 \pi l_B^2 \sigma z_{\rm s}^3 $. \\
The excluded volume correction to the electrostatic potential of the ions can be approximated by $\beta e \phi^{\rm ex}=\frac{1}{2}\int_0^{d_{ii}} \frac{l_B}{r} 2 \pi r \sigma \, dr=\pi l_B d_{ii} \sigma$, where $d_{ii}$ is the closest separation between two ions, and is similar in spirit as the hole corrected Debye-H\"uckel theory \cite{Nordholm84}.
This term, being linear in $\sigma$, resembles a parallel capacitor correction. \\
The only undefined quantity is now the $\phi^{\rm mf}$, but this quantity is redundant as it appears in both Eqs.~\ref{contact} and \ref{titr}. The surface charge density and $z_{\rm s}$ can, hence, be solved self-consistently \footnote{$\Xi$ and $z_{\rm s}$ need both to be updated at each iteration step.}.  \\
\newline
{\it Results -} We study the same systems as in Refs.~\cite{Labbez09} and \cite{Dove05}, with $l_B=0.714$ nm corresponding to pure water at room temperature, $\sigma^{\rm site}=4. 8$ sites/nm$^2$ (corresponding to a maximum surface charge density of -770 mC/m$^2$), p$K_a=7.7$, $d_{is}=0.35$ nm, and $d_{ss}=0.4$ nm.  These values give $C_{\rm S}=1.93$  F/m$^2$, which is lower than the value $C_{\rm S}=2.6$ F/m$^2$ used in Ref.~\cite{Labbez09} to fit the surface complexation model with their MC results for simple 1:1 salts. However, combining the Stern capacitor with the ``excluded volume'' capacitor gives $C_{\rm eff}^{-1}=C_{\rm S}^{-1}+C_{\rm ex}^{-1}=2.67$ F/m$^2$, a value very close to the empirical fitted value of Ref.~\cite{Labbez09}.
This gives a theoretical sound explanation of the increased capacitance needed in mean-field theories to agree with MC results. Furthermore, two kinds of simulation,, which we will compare to, were carried out in Ref.~\cite{Labbez09}, a set of simulations with discrete surface site charges and the other with uniformly smeared out surface site charges.
For more information about the MC simulations, see Ref.~\cite{Labbez09}.\\
Fig.~\ref{fgr:Titr11} shows the effect on surface ionisation as a function of p$H$ 
of having (a) uniformly smeared site charges compared to (b) accounting for their discreteness for both the MC simulations and present theory at various 1:1 salt concentrations.
We see that the surface charge density, both in the simulations and the theory, is increased once we account explicitly for the discreteness of the surface site charges. The MC and theory results show such a remarkable agreement that the two different lines are indistinguishable for a given 1:1 salt concentration, both of which agree with the experimental curves.  
Including also ion-ion correlations into the theory only alters the titration curves by at most $\sim$5\%, at the highest studied p$H$-values, as the average coupling parameter is too weak to influence the results (see SI). 
Looking at the inset of Fig.~\ref{fgr:Titr11}(b), one finds that the site-site correlations increase the surface charge densities at all studied p$H$ values, with a relative increase from $\sim$5\% at p$H$ of 4 to $\sim$30\% at the highest studied pH value. However, at lower p$H$ values, the surface charge densities are so low that the increase typically is within the noise of the experimental data or MC results. Nevertheless, combining Eqs.~\ref{titr} and \ref{eq:ss}, we find that the surface charge density increases by
\begin{equation}
\sigma/\sigma^{\rm \bf(a)}\simeq 1+M_{\rm sq} l_B \sqrt{|\sigma|}
\label{eq:lowI}
\end{equation}
at low surface ionisations when site-site correlations are accounted for compared to without (the latter denoted by $\sigma^{\rm \bf(a)}$, \emph{i.e.}~the lines shown in Fig.~\ref{fgr:Titr11}(a)) at the same salt and p$H$ conditions. \\
We continue testing the theory for 2:1 salts. Fig.~\ref{fgr:Titr21} compares the new approach, including various correlation terms, with experimental and MC results. Notice that the 67 mM CaCl$_2$ and 200 mM NaCl
have the same Debye screening length and, hence, the linearised Poisson-Boltzmann (\emph{i.e.},~Debye-H\"uckel) would predict the same titration curve for the two salt cases.
The difference between the two corresponding theory lines in Fig.~\ref{fgr:Titr21}(a), without any correlations, is therefore due to the non-linearity of the Poisson-Boltzmann equation, which starts to be significant at high surface potentials and high ion valencies.
Fig.~\ref{fgr:Titr21}(b) shows that predictions can be improved by including only the site-site correlations up to a surface charge density of $\sigma \sim 100$ mC/m$^2$ for the 2:1 salt.
Including also ion-ion correlation effects into the theory improves the agreement for higher surface charge densities, as seen from Fig.~\ref{fgr:Titr21}(d), even though it is now visible that the MC simulations and the theory do not render exactly the same values in the high p$H$ regime, where the surface charge densities are high. This finding contrasts with the 1:1 salt cases, where we found that ion-ion correlations are almost negligible for all the studied surface charge densities (see SI). As the divalent ions act as counterions to the surface, the corresponding
coupling parameter $\Xi\sim \sigma z_s^3$  is roughly a factor 8 bigger than the 1:1 salt case at a comparable surface charge density, explaining the importance of ion-ion correlations in the former but not the latter.\\
There might be several reasons for the minor discrepancy between MC and the theory seen at high surface charge densities and for divalent counterions: having approximate ion-ion and site-site correlation terms, ignoring hardcore effects or neglecting the ion-site correlations.  We set out to try one of these hypotheses: the approximate ion-ion correlation term.  
Fig.~\ref{fgr:Titr21}(c) compares our theory to MC simulations with smeared-out surface charge density. These structureless surfaces have zero site-site and ion-site correlations and serve as an excellent way to test the ion-ion correlation approximation. The comparison between the MC results and the theory, including ion-ion correlations, shows that the ion-ion correlations approximation is a very good, if not an excellent, approximation. \\
Including both ion-ion and site-site correlations into our theory overestimates the surface charge density by roughly $\sim 15\%$ at the highest p$H$ values. This overestimation indicates that the ion-site is destructive compared to the two other correlation terms. 
We can continue and assume that the main consequence of ion-site correlations is a partly neutralising effect of surface charges of the sites. We, hence, postulate that this ion-site potential
can be expressed as:
\begin{equation}
\beta e\phi^{\rm is} \simeq  \mathrm{sgn}(\sigma) \Delta M_{\rm sq} l_B \sqrt{|\sigma|},
\end{equation}
with $\Delta M_{\rm sq}=c_1 \Xi$. This term, hence, partly cancels the $\phi^{ss}$ term as would be expected if the surface site charges are partly neutralised, with the neutralisation strength proportional to the effective counterion valency. Assuming that the ions at the surface neutralise the surface charge density, the proportion of ions with an average charge of $z_{\rm s}$ is inversely proportional to $z_{\rm s}$. The energy gain/loss by incorporating the ion-site correlations, including corrections both to the ions and sites, in this way is close to zero, \emph{i.e.},~$-\beta e\phi^{\rm is}+z_{\rm s} \beta e\phi^{\rm is}/z_s=0$, and should be regarded as a conservative estimate of the ion-site correlations. We can empirically improve our data with this approach by using $c_1=0.029$, see Fig.~\ref{fgr:Titr21F}. \\
The Monte Carlo study \cite{Labbez09} assumed that the difference between smeared out surface charge density and wall with discrete surface charges was due to the lack of ion-site correlations in the former case. Here we show that this is predominantly due to site-site correlations in the 1:1 salt case and a combination of site-site and ion-ion correlations for the 2:1 salt case. Ion-site correlations have only a marginal effect on the titration curves for 1:1 salts ($<5$\%) and only mildly ($<15$\%) for the 2:1 salt case. Previously studies for counterion only systems have reached the same conclusion that ion-site correlations primarily are negligible \cite{Moreira02}. \\ 
We can gain further insights into the various contributions by looking at the relative increase of the different terms, as shown as insets in Fig.~\ref{fgr:Titr21} and \ref{fgr:Titr21F}. For a 2:1 salt, as for the 1:1 salt case, the site-site correlations are always at play (even down to a p$H$ of 4, see insets of Figs.~\ref{fgr:Titr21}(b) and (d)), while  the ion-ion correlations start to be relevant first above a p$H$ of 6-7, \emph{i.e.,} slightly below p$K_a$, (see insets of Figs.~\ref{fgr:Titr21}(c) and (d)), in line with the observations of Labbez \emph{et al.}~\cite{Labbez09}. Ion-site correlations are found only to mildly affect the titration curve at high p$H$-values, which is evident by the different scale used in the inset of Fig.~\ref{fgr:Titr21F} compared to the insets of Fig.~\ref{fgr:Titr21}. Notice that adding the correlation terms increases the overall surface charge density up to a factor of 2 compared to the most straightforward approach without any correlations included.\\
Labbez \emph{et al.}~\cite{Labbez09} also explored the effect of surface charge density as two surfaces approach each other and found that the curve was non-monotonic for 2:1 salts and high p$H$ values.
For finite distances, one has a finite pressure $\beta P$. According to the contact theorem, this non-zero pressure leads to a correction term on the right side of Eq.~\ref{contact} as 
\begin{equation}
2 \pi l_B \sigma^2 = \sum_i (e^{-\beta z_i e \phi^{\rm ion}}-1)\rho_{i}^{\rm bulk}-\beta P.
\label{contact2}
\end{equation}
The charge regulating (CR) condition is typically between the limiting conditions of constant charge (CC) and constant potential (CP) \cite{Markovich16}. By definition, the CC condition yields constant  surface charge densities for all separations. For the CP condition, however, the surface charge densities would depend on the pressure. For a 1:1 salt, the pressure increases monotonically as the separation decreases; hence the surface charge density decreases monotonically as a function of decreasing separation. However, for a 2:1 salt, there exist regimes where the pressure is attractive due to ion-ion correlations \cite{Guldbrand84}. At these separations, the surface charge density would instead increase, according to Eq.~\ref{contact2}. CR being a mixture of CC and CP, one quickly sees that the CR condition has the same qualitative behaviour as the CP condition, \emph{i.e.,} the surface charge densities follow the pressure curve trend of the CP condition. Hence, we can understand the increasing surface charge densities as a function of decreasing separation for 2:1 salts and high p$H$ values due to the presence of attractive ion-ion correlation mediated surface-surface forces in line with the conclusions of Ref.~\cite{Labbez09}. \\
\newline
{\it Summary -} We have presented a new and simple theory for the titration behaviour of charged planar surfaces, including site and ion correlations. Our approach is in excellent agreement compared both to experiments and Monte Carlo simulations. The results show that site-site correlations always are at play, being one of the dominant correction terms. In contrast, the ion-ion correlations are first significant for divalent salts and high surface charge densities, \emph{e.g.,} high effective coupling strengths. We find that ion-site correlations are, for the most, subdominant. \\
This work generalises the strong coupling approach for salts, relying on an effective coupling parameter, different and complementary to the dressed ion approach \cite{Kanduc11}. Further work will include excluded volume effects, ion-specific effects, and detailed investigations of the density profiles close to the surfaces \cite{Bakhshandeh11,Bakhshandeh19}. 
\\
\newline
The author thanks C. Labbez for providing the data and both C. Labbez and M. Ullner for valuable comments.

\section{Supplementary Information}

\setcounter{figure}{0}
\makeatletter 
\renewcommand{\thefigure}{S\@arabic\c@figure}
\makeatother

\begin{figure}[!htbp]
\centering
\includegraphics[width=0.23\textwidth]{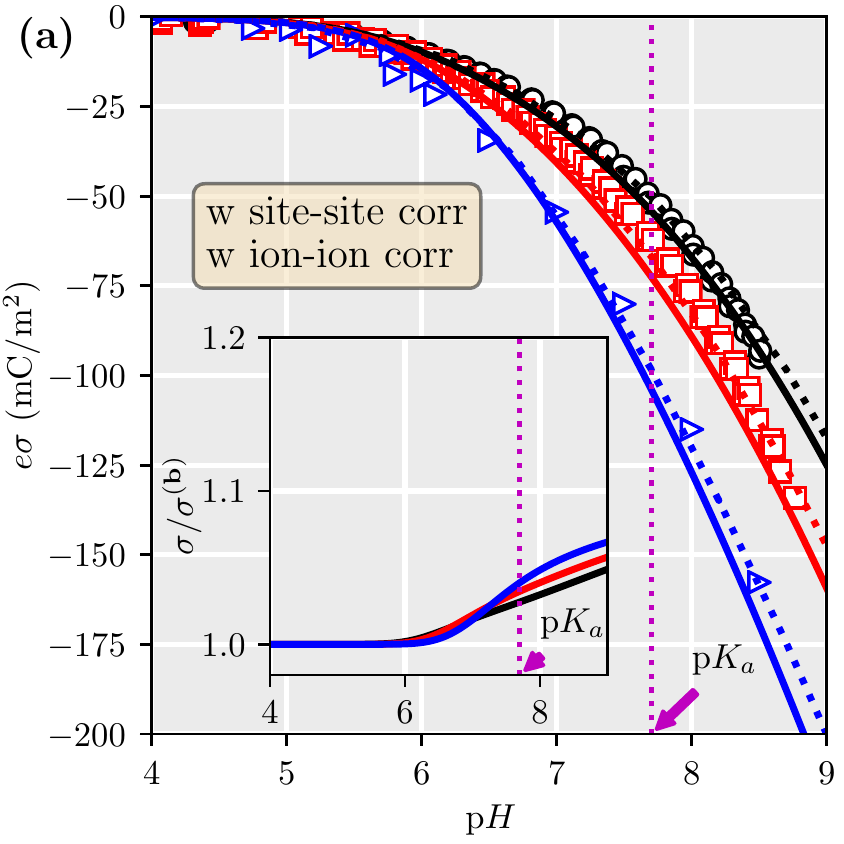}
\includegraphics[width=0.23\textwidth]{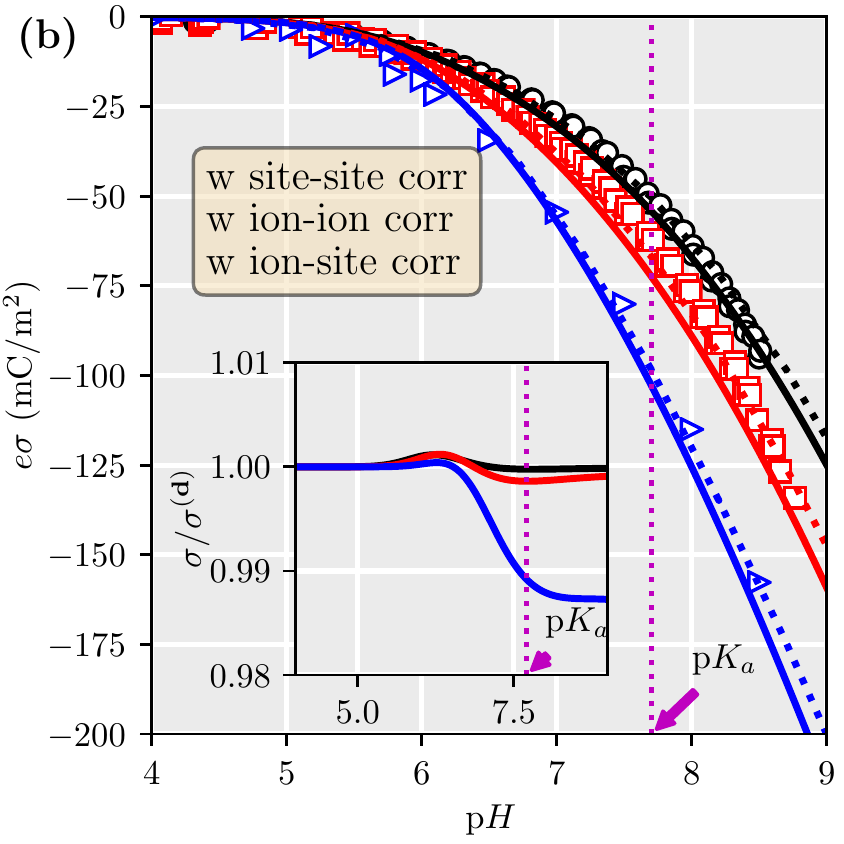}
  \caption{Effect on the titration curve for 1:1 salt, including, besides site-site correlations, also {\bf (a)} ion-ion correlations and {\bf (b)} all ion correlations. Inset in {\bf (a)} shows the relative shift in the surface charge density compared to only including site-site correlations, and the inset in {\bf (b)} the relative shift accounting for ion-site correlations compared to without. }
 \label{fgr:Titr11c}
\end{figure}

Fig.~\ref{fgr:Titr11c} shows the effect, including ion-ion correlation and ion-site correlations, on the titration behaviour as a function of p$H$ at various concentrations of 1:1 salts. The is a slight increase in surface charge density at the same conditions (\emph{i.e.}, salt concentration and p$H$), leading to a slightly worse agreement with the Monte Carlo results. However, the error is less than $\sim$5\%, found at the highest p$H$ values, and given the approximative theory, this is still a tolerable error. 

\end{document}